

\magnification=1200
\hoffset=-.1in
\voffset=-.2in
\vsize=7.5in
\hsize=5.6in
\tolerance 500

\def\ll{\left\langle}
\def\rr{\right\rangle}
\def\rapp{\hbox{${\lower.40ex\hbox{$>$}\atop \raise.20ex\hbox{$\sim$}}$}}
\def\thru#1{\mathrel{\mathop{#1\!\!\!/}}}
\def\Tr{\,{\rm Tr}\,}
\def\footnoterule{\kern-3pt \hrule width \hsize \kern2.6pt}

\pageno=0
\footline={\ifnum\pageno>0 \hss --\folio-- \hss \else\fi}

\baselineskip 15pt plus 1pt
\centerline{\bf HEAVY QUARK FRAGMENTATION}
\smallskip
\centerline{{\bf INTO HEAVY MESONS}%
\footnote{*}{This work is supported in part by funds
provided by the U. S. Department of Energy}%
\footnote{}{(D.O.E.) under contract \#DE-AC02-76ER03069,
and in part by the Texas National}%
\footnote{}{Research Laboratory Commission under grant \#RGFY92C6.}}
\vskip 24pt
\centerline{R.~L.~Jaffe and L.~Randall\footnote{$^\dagger$}{National
Science Foundation Young Investigator Award.}\footnote{}{Alfred
P.~Sloan Foundation Research Fellowship.}\footnote{}
{Department of Energy Outstanding Junior Investigator Award.}}
\vskip 12pt
\centerline{\it Center for Theoretical Physics}
\centerline{\it Laboratory for Nuclear Science}
\centerline{\it and Department of Physics}
\centerline{\it Massachusetts Institute of Technology}
\centerline{\it Cambridge, Massachusetts\ \ 02139\ \ \ U.S.A.}
\vfill
\centerline{\bf ABSTRACT}
\baselineskip = 18pt plus 1pt
\medskip
We present a QCD based interpretation of heavy quark fragmentation
which utilizes the heavy quark mass expansion. By distinguishing
between perturbative and non-perturbative QCD effects, we show how
to reliably extract mass independent parameters characterizing
the fragmentation function.  Because these parameters are
quark mass independent,
 this procedure should permit tests of heavy
quark symmetry.  Furthermore,
we show that heavy quark mass corrections
vanish at order $m^2/Q^2$
in QCD.  There also exist higher twist
corrections of order $\Lambda m/Q^2$  and ${\alpha_{QCD}\over\pi}{m^2\over
Q^2}\ln (Q^2/m^2)$
which we relate to the
leading twist fragmentation function.
\vfill
\centerline{Submitted to: {\it Nuclear Physics B\/}}
\vfill
\noindent CTP\#2189 \hfill hep-ph/9306201 \hfill May 1993
\eject
\baselineskip 24pt plus 2pt minus 2pt
\noindent{\bf I.\quad INTRODUCTION}
\medskip
\nobreak
Heavy quark effective theories have led to considerable insight into the
properties of hadrons containing $b$- and $c$-quarks.  Key among the
assumptions underlying the heavy quark theory is that one can
expand matrix elements in $\Lambda/m$,
where $\Lambda$ is some QCD related scale, $m$ is the heavy quark mass, and
where coefficients in the expansion are independent of $m$.  We show in this
paper that a similar expansion of the moments of the heavy quark fragmentation
function leads to valuable information about the heavy quark fragmentation
function both in leading twist and beyond. We derive a parameterization of the
fragmentation function which involves heavy quark mass independent parameters
and show how the leading parameters
 can be readily extracted from the lowest moments.
This permits
tests of heavy quark symmetry.  In particular,
the fragmentation function of the $b$- and $c$-quarks, both of which
have been already studied and will be better measured in the future,$^{1,2}$
are readily related.
We show that the commonly used Peterson fragmentation function$^3$
has a misleading dependence on the QCD scale and heavy quark mass;
the width is linear and not quadratic in the ratio.

The basic idea underlying
 our analysis is very simple. Fragmentation and distribution
functions measured at a scale $Q^2$ evolve from ``boundary data'' ---
forward matrix elements of bilocal heavy quark operators --- supplied at
some
reference scale $\mu^2$.  This boundary data incorporates the
non-perturbative
features of the fragmentation function (at least in regions
sufficiently far
from
$z=0$ and
$z=1$ where
we will see
heavy quark methods and perturbation theory respectively
break down). By evaluating the matrix
elements at  sufficiently low renormalization mass scales,
one can exploit heavy
quark symmetry to derive a heavy quark independent parameterization
of the structure function.
At this low mass scale,
the boundary data can be organized as a power series in
$\Lambda/m$, with mass independent coefficients. The
measured fragmentation function is then obtained by
  evolving the function to $Q^2$
via perturbative QCD.

Previous work on heavy quark fragmentation functions either has been
completely phenomenological$^3$ or has
concentrated on the fragmentation
function generated by perturbative evolution alone.$^4$
 Some authors have explored the possibility of describing
 the fragmentation function solely in terms of
perturbative QCD.$^5$  We abandon the objective of previous authors of
deriving the fragmentation function.  We distinguish between perturbative
contributions which can be reliably calculated at high momenta and
non-perturbative contributions which we parameterize (but don't calculate) by
exploiting the heavy quark mass expansion.
We show
 that the expansion of the moments of
the fragmentation function at any given order
in $\Lambda/m$ only involves a small number of
mass independent parameters.
Because the same parameters enter the $b$ and $c$ quark fragmentation
functions and moments of different order, tests of heavy quark symmetry will be
possible.

The principle results of our analysis
are:
\medskip
\item{$\bullet$}At leading twist, we find a simple parametric form for the
quark mass dependence of the fragmentation and distribution functions at the
heavy quark mass scale which applies at leading order in $\Lambda/m$.  Our
form differs from the commonly used function of Ref.~[3] because
our prediction for the width is order $\Lambda/m$, and not order
$\Lambda^2/m^2$.  This difference is critical to extracting
the scale $\Lambda$ which enters the heavy quark expansion.
\medskip
\item{$\bullet$}We show that the moments of the fragmentation function can be
interpreted in terms of mass-independent parameters.  We
concentrate
on only a finite number of these parameters
which we expect  can be reliably extracted.
This
will allow for  tests of heavy quark symmetry. Moreover, it could
provide further data on which scale $\Lambda$ works best
in the heavy quark expansion.
\medskip
\item{$\bullet$}Within our framework it is possible to make rather
strong statements about
those higher-twist effects proportional to powers of $m^2/Q^2$, where $Q^2$
is the center-of-mass energy squared.  Since $m>\!\!>\Lambda$, these are the
most important higher-twist corrections.  We show that the sum of all
corrections of order $\left( m^2/Q^2\right)^k$ to the process $e^+e^-\to H(P)+
X$ vanishes for all $k>0$ at leading order in $\alpha_s$. Radiative corrections
modify this result and give rise to corrections of order ${1\over\pi} \alpha_s
(Q^2) {m^2\over Q^2}\ln {Q^2\over m^2}$.  We show that these corrections can be
related back to the leading twist fragmentation function.  Mass suppressed
corrections are of order $\Lambda m/Q^2$ and can also be related back to the
leading twist fragmentation function.
\medskip
The paper is organized as follows.  In Section~II we derive the leading twist
results, first for distribution functions, which are pedagogically simpler,
although not the case of greatest experimental interest, and then for
fragmentation functions.   In Section III, we investigate the applications
of our approach. We conclude that a finite number of non-perturbative
parameters can be reliably extracted
by measuring the moments of the fragmentation
function, using
a simple phenomenological form whose parameters are readily interpreted in a
heavy quark mass expansion.
In Section~IV we consider higher-twist contributions to
$e^+e^-\to H(P)+X$, after which we conclude.
\goodbreak
\bigskip
\vfill
\eject
\hangindent=23pt\hangafter=1
\noindent{\bf II.\quad HEAVY QUARK DISTRIBUTION AND FRAGMENTATION \hfil\break
FUNCTIONS AT LEADING TWIST}
\medskip
\nobreak
\noindent{\bf II.1\quad Distribution Functions}
\medskip
\nobreak
Figure 1 shows the virtual Compton scattering
diagram for an $h$ quark current whose imaginary part yields the dominant
contribution to a hypothetical
deep inelastic scattering experiment from an $H$-meson target.
A standard analysis of this diagram yields
the distribution function for heavy quarks ($h$ with mass
$m$) in a heavy meson ($H$ with mass $M$)
in terms of a
Fourier transform of a heavy quark correlation function along the null
plane:$^6$
$$f\left( x,\mu^2\right) = {1\over 4\pi}\int d\lambda\,e^{i\lambda x} \ll H(P)
\left| \overline{h} (0) \thru{n} h (\lambda n) \right|H (P) \rr_{\mu^2}
\ \ .\eqno(2.1)$$
Here $n^\mu$ is a null vector defined by $P^\mu = p^\mu + M^2 n^\mu\big/ 2$,
the heavy meson momentum and $p$
and $n$ are defined by $p^2 = n^2 = 0$, $p\cdot n=1$ yielding $p^\mu =
p(1,0,0,1)$, $n^\mu = (1/2p)(1,0,0,-1)$. The variable $x$ is related to the
kinematic variables of Compton scattering by $x~=~Q^2/{2q\cdot P}$.
The parameter $p$ is arbitrary and fixes the frame.  For example, $p=M/2$
corresponds to the heavy meson rest frame; $p\to\infty$ is the ``infinite
momentum frame'' often used in parton models.  Here, $h(\xi)$ is the heavy
quark field.   Equation~(2.1) is written in $n\cdot A = 0$ gauge, avoiding
the need for a line integral between $\overline{h}$ and $h$. Also we have
suppressed a factor of $n \cdot P=1$ in the exponent. Contraction with
$n^\mu$ is natural in deep inelastic processes in which some direction of
hard momentum flow is specified ({\it e.g.\/} $q^\mu\cong \nu n^\mu -
xp^\mu$ in deep inelastic scattering).  It also selects operators of pure
twist (twist-2 in this case) since traces are removed by contraction with
$n^\mu$. The mass scale $\mu^2$ is the
renormalization scale required to render the local operators in the Taylor
expansion of (2.1) finite.  Equivalently, $\mu^2$ appears as a cutoff on the
transverse momentum integral which arises in a momentum space representation
of $f(x,\mu^2)$.

In order to apply heavy quark symmetry we should take
$\mu^2=\mu_0^2$, where $\mu_0$  is
a scale beneath the lightest of the heavy quark masses,
but above the QCD scale so that perturbative QCD is still reliable.
However, we will see later that to the order of accuracy we can hope to
achieve, our results apply at any renormalization scale below $m^2$.  Thus
we will eventually set $\mu_0^2=m^2$, but for now we keep it as an
independent renormalization scale. The scale $m^2$ is time-like whereas $\mu^2$
is space-like.  This distinction
only affects terms higher order in $\alpha_s(\mu^2)$ than concern us here,
so we drop it.

It is straightforward to convert $f(x,\mu_0^2)$ to a form in which its
interpretation as a parton momentum distribution is more readily apparent.
First, decompose $h$ with the light-cone projection operator $P_\pm = {1\over
2} \gamma^\pm \gamma^\mp$ with $h_+=P_+h$, and insert a complete set of
intermediate states $\{|\chi\rangle\}$,
$$f(x,\mu_0^2) = {1\over 2p} \sum\limits_\chi \delta\left( x - 1 + P_\chi
\cdot
n\right)\left| \ll \chi\left| h_+ (0) \right|H(P)\rr\right|^2 \eqno
$$
where we have suppressed reference to the renormalization scale on the
right-hand side.  The momentum $P^\mu_h = P^\mu - P^\mu_\chi$ is the momentum
of the struck heavy quark.  The function $f(x)$ measures the probability of
finding a ``good'' light-cone component of the quark field with momentum
fraction $x\left( p^+ = xP^+\right)$ in the target.  The physical process
$eH\to eX$ also receives contributions from the heavy {\it antiquark\/}
distribution in $H$,  $\overline{f}(x,\mu^2)$, corresponding
to Figure 1 with the quark line replaced by an antiquark line.
It is difficult to separate $\overline{f}$ from $f$ experimentally.
The function  $\overline{f}(x,\mu^2)$ is defined by the crossing
relation $
\overline{f} (x,\mu^2) = - f(-x,\mu^2)$ and has a parton representation
like (2.2) with $h_+$ replaced by $h^\dagger_+$.  It measures the probability
of finding a heavy antiquark in the meson $H$.    We argue below that
$\overline{f}(x,\mu^2)$ is
 small at  $\mu^2\sim m^2$.  Similarly, a potential gluon
contribution is also small  at $\mu^2\sim m^2$.

We now proceed to parameterize (2.1).  We first expand
$$f(x,\mu_0^2) = {1\over 4\pi} \sum^\infty_{k=0} {1\over k!} \int
d\lambda\,e^{i\lambda x} (-i\lambda)^k\ll H(P)\left|\overline{h}(0) n_\alpha
\gamma^\alpha (n\cdot\Pi)^k h (0) \right| H(P)\rr \eqno(2.3)$$
where $\Pi_\mu = i D_\mu$ and we have returned to a manifestly gauge
invariant form.
We match this matrix element of the full theory onto
matrix elements in the heavy quark effective theory.
We define the meson four velocity,
$v^\mu$, by $P^\mu=Mv^\mu$, where $M$ is the meson mass.
Following the standard procedure, (although we do
not bother to rescale the fields, since everything will
be dimensionless in the end)
we decompose the field $h(\xi)$ as
$$h(\xi) = h_v(\xi) + \underline{h}_v(\xi) \eqno(2.4)$$
where
$$\eqalignno{h_v(\xi) &= e^{-imv\cdot\xi} {\cal P}_+h(\xi) &(2.5) \cr
\underline{h}_v(\xi) &= e^{-imv\cdot\xi}{\cal P}_- h(\xi) &(2.6) \cr}$$
where ${\cal P}_\pm = (1\pm\thru{v})/2$.  Then
$$\Pi_\mu h (\xi) = e^{-imv\cdot\xi} \left( mv_\mu + \Pi_\mu\right) \left(
{\cal P}_+h(\xi) + {\cal P}_-h(\xi)\right) \ \ .\eqno(2.7)$$

We now match the matrix element which appears in (2.3),
$$\Gamma_k = \ll H(P) \left|\overline{h} (0)
n_\alpha \gamma^\alpha(n\cdot\Pi)^k
h(0)\right|H(P)\rr \eqno(2.8)$$
onto a sum of matrix elements of the heavy quark effective theory.
Clearly, these  matrix elements can be organized as a power series
in  $1/m$.  The leading term is the term in which the momentum is $m v_\mu$
and the matrix element is taken between the fields $h_v$.
Mass suppression
factors in the remaining matrix elements
derive from several sources.
First, in the
derivative in (2.7), the first term is
 $mv_\mu h_v(\xi)={m\over M}P_\mu h_v(\xi)$, while the second (mass suppressed)
term is  $\Pi_\mu
h_v(\xi)\approx \epsilon P_\mu h_v(\xi)$ where $\epsilon \approx 1 - {m\over M}
\equiv
{\overline{\Lambda}\over M}$.  Second, because the equation of motion yields
$\underline{h}_v(\xi) = (1/2m)\Pi^\alpha \gamma_\alpha h_v(\xi)$, the matrix
elements involving $\underline{h}_v(\xi)$ are mass suppressed relative to
matrix elements of the same operator with $h_v(\xi)$ instead.  Finally, matrix
elements  involving
the time-ordered product of the operator with mass suppressed operators,
${\cal
O}_{1/m}$, in the heavy quark Lagrangian also yield mass suppression.  With
these considerations in mind, we substitute (2.7) into (2.8) and organize the
expansion according to the powers of $1/m$,
$$\eqalign{
\Gamma_k &= \sum\limits_\ell \Biggl\{ \left(\matrix{k\cr\ell\cr}\right)
\left( n \cdot mv\right)^\ell \ll H(P) \left| \overline{h}_v (0) n_\alpha
\gamma^\alpha \left( n\cdot\Pi\right)^{k-\ell} h_v (0) \right| H(P)\rr \cr
&+ \left( \matrix{k\cr\ell+1\cr}\right) \left( n\cdot mv\right)^{\ell+1}
\Biggl[ \ll H(P) \left|
\overline{\underline{h}}_v(0) n_\alpha \gamma^\alpha \left(
n\cdot\Pi\right)^{k-\ell-1} h_v(0)\right| H(P)\rr \cr
&+ \ll H(P) \left| \overline{h}_v(0)n_\alpha\gamma^\alpha \left( n\cdot
\Pi\right)^{k-\ell-1} \underline{h}_v(0)\right| H(P)\rr \cr
&+ \ll H (P) \left| T\left( {\cal O}_{1/m} \overline{h}_v(0)
n_\alpha\gamma^\alpha \left( n\cdot\Pi\right)^{k-\ell-1} h_v(0)\right)
 \right| H(P)\rr\Biggr] + \ldots \Biggr\}\cr} \eqno(2.9)$$
where $\ldots$ represents terms multiplying
$\left(\matrix{k\cr\ell+n\cr}\right)$, $n>1$.  In (2.9),
$\left(\matrix{k\cr\ell\cr}\right)$ is the binomial coefficient defined to be
zero for $\ell<0$ or $\ell>k$.  The range of the sum on $\ell$ is then
dictated by the values for which $\left(\matrix{k\cr\ell\cr}\right)\not=0$.
The first term has the largest binomial coefficient and comes from
terms in which all the mass suppression factors come from the
small momentum piece of $\Pi_\mu$.  In the second term,
one power of mass suppression has a different origin.  The $\ldots$
represents terms with even smaller binomial coefficient,
which begin at order $1/m^2$.

We now  rewrite the above as
$$\eqalign{M_k &= 2\sum\limits_\ell\left(\matrix{k\cr\ell\cr}\right) \left(
n\cdot mv\right)^\ell \epsilon^{k-\ell} A_{k-\ell} \cr
&+ \left( \matrix{k\cr\ell+1\cr}\right) \left( n\cdot mv\right)^{\ell+1}
\epsilon^{k-\ell} B_{k-\ell} \cr
&+ \left( \matrix{k\cr\ell+2\cr}\right) \left( n\cdot
mv\right)^{\ell+2}\epsilon^{k-\ell} C_{k-\ell}+\ldots \cr} \eqno(2.10)$$
where
$$\ll H(P) \left|\overline{h}_v(0)
n_\alpha \gamma^\alpha \left( n\cdot\Pi\right)^k
h_v(0) \right| H(P) \rr = 2A_k \epsilon^k \eqno(2.11)$$
and similarly for $B$ and $C$, defined in terms of the sum of matrix elements
multiplying a particular binomial coefficient.
\footnote{*}{Although order $\epsilon$ terms can
be shown to vanish, this will not be the
case for the fragmentation function. We simply
leave in all the terms here. }
All dependence on the heavy quark mass is contained in the factors
$\epsilon^p$ provided $\epsilon$ scales
inversely to the heavy meson mass. We make the specific choice
$\epsilon=\overline{\Lambda}/M = 1-m/M$ which simplifies
the algebra.  Notice that $\epsilon$ scales like the
inverse heavy {\it meson\/} mass because  $n\cdot P=1$.  To
get  the
heavy {\it quark\/} mass expansion in the end
requires expanding the meson mass $M$ in terms
of the quark mass $m$.
Furthermore, because  the heavy quark
occurs in the functions $B$, $C$, $\ldots$, one
must be careful to expand the mass factor here in terms
of the meson mass for consistency. With only the functions
$A$, $B$, and $C$, the expansion above is sufficient.

 Having extracted the factors of $\epsilon$, the
parameters $A$, $B$ and $C$ are independent of the heavy quark mass.  Here,
$A_0=1$ because $\overline{h}\thru n h$ counts heavy quarks minus
antiquarks
in the $H$-meson.
\goodbreak
At this point, we wish to exploit the fact that the same matrix elements
appear in all the $\Gamma_k$.  We will first add the terms into a general
function incorporating this fact.
Following this, we will derive relations among
moments.  It is useful to define functions $a(y)$, $b(y)$, {\it etc.\/} whose
moments give the $A_k$, $B_k$, {\it etc\/}.  We define
$$\eqalignno{A_k &= \int^1_{-\infty} dy\, a(y) y^k &(2.12) \cr
B_{k+1} &= \int^1_{-\infty} dy\, b(y)y^k\ \ .&(2.13)\cr}$$
The limits on the $y$-integration are chosen to reproduce the required support
of the structure function $(0<x\le 1$).  Note that $a(y)$ and $b(y)$ must
fall rapidly as $y\to -\infty$ in order that (2.12) and (2.13) converge.
The exact nature of this requirement and its physical significance will be
further discussed below.
  The upper
limit is unity due to our choice of $\epsilon$. So long as $\epsilon$ scales
like $1/M$ the upper limit is mass independent.    We now have
$$\eqalign{\Gamma_k &= 2 \sum_\ell\int^1_{-\infty}
dy \left[\left(\matrix{k\cr\ell\cr}\right)\left(
{m\over M}\right)^\ell \left(\epsilon y\right)^{k-\ell} a(y) + \right.\cr
&\left. \left(
\matrix{k\cr\ell+1\cr}\right) \left( {m\over M}\right)^{\ell+1}
\epsilon\left(\epsilon y\right)^{k-\ell-1}b(y) + \ldots \right] \cr
&= 2\int^1_{-\infty}dy \left[\left( {m\over M} + \epsilon y\right)^k a(y) +
\epsilon \left( {m\over M} + \epsilon y\right)^k b(y) + \ldots\right]
\cr}\eqno(2.14)$$
where we have used $n\cdot mv=m/M$.  We can now substitute
$\Gamma_k$ into (2.3) to obtain
$$\eqalignno{
f(x,\mu_0^2) &= \sum_k \int^\infty_{-\infty}
{d\lambda\over{2\pi}} \int^1_{-\infty}dy\,
e^{i\lambda
x} {1\over k!} \left( -i\lambda\right)^k \left( {m\over M} + \epsilon
y\right)^k \left( a (y) + \epsilon b(y) + \ldots\right)
 \qquad&\qquad(2.15) \cr
&= {1\over 2\pi}\int^1_{-\infty}dy\int^\infty_{-\infty}\,
d\lambda\,e^{i\lambda\left( x-m/M-\epsilon
y\right)} \left( a(y) + \epsilon b(y) + \ldots \right) &(2.16) \cr
&= \int^1_{-\infty} \delta\left( x - m/M - \epsilon y\right)\left( a(y) +
\epsilon b(y) + \ldots\right) &(2.17) \cr
f(x,\mu_0^2) &= \left[ {1\over\epsilon}a \left( {x-{\displaystyle{m\over M}}
\over
\epsilon}\right) + b \left( {x-{\displaystyle{m\over M}}\over\epsilon}
\right) + \ldots \right] &(2.18) \cr}$$

Equation~(2.18) is the result from which our conclusions about the heavy
quark distribution function follow.  It expresses the fact that $f(x,\mu_0^2)$
can be expanded in a power series in
$\epsilon=\overline{\Lambda}/M$.  At any order in $\epsilon$, there is
a single function.  However, the argument of that function,
$$y = {x-{\displaystyle{m\over M}}\over\epsilon} = {x-{\displaystyle{m\over
M}} \over 1 - {\displaystyle{m\over M}}}\ \ ,\eqno(2.19)$$
depends in a rather singular way upon $\epsilon$.

The derivation of (2.18) suggests that it holds only if the functions
$a(y)$, $b(y)$, {\it etc.\/} fall faster than any power of $y$ as $y\to
-\infty$ as indicated by (2.12) and (2.13).  In fact this
assumption is stronger than is required.  It
was convenient for us to expand the distribution function in terms of
local operators and then resum into $f(x)$.  Had we instead matched
directly onto the bilocal operator product of (2.1),
we would have achieved (2.18) directly without any requirement that {\it
all\/} of the $A_k$, $B_k$, {\it etc.\/} exist.  The steps in such a
derivation are straightforward, so we summarize them very briefly here.
First, perform the matching summarized by (2.5) and (2.6) directly in (2.1).
The result is

$$f\left( x,\mu^2\right) = {1\over 4\pi}\int d\lambda\,e^{i\lambda x} \ll H(P)
\left| \overline{{\cal P}_+h} (0) \thru{n} {\cal P}_+h (\lambda n)
\right|H (P) \rr\,e^{-im\lambda v\cdot n} + \ldots
\ \ ,\eqno(2.20)$$
where we have kept only the term which will eventually turn into $a(y)$
and  $\ldots$ denotes all other terms. Eq.(2.20) remains in $n\cdot A = 0$
gauge (but can be made gauge invariant by restoing the line integral)
and we suppress the renormalization scale $\mu^2$.  The matrix
element in (2.20) is dimensionless, remembering the dimension of $n_{\mu}$, so
converting to the normalization of the heavy quark theory it turns into a
dimensionless function of the variable $\lambda{\Lambda\over M}$,
$$\ll H(P)\left| \overline{{\cal P}_+h} (0) \thru{n}
{\cal P}_+h (\lambda n) \right|H (P) \rr = {\cal
F}\left(\lambda\displaystyle{\Lambda\over M}\right)\ \ , \eqno(2.21)$$
which we represent as a Fourier transform,
$${\cal F}\left(\lambda\displaystyle{\Lambda\over M}\right) =
2\int_{-\infty}^\infty d\alpha\,e^{-i\alpha\lambda{\Lambda\over
M}}a(\alpha)\ \ . \eqno(2.22)$$
It is then a simple matter to substitute this Fourier representation into
(2.20) and obtain (2.18).  It is clear from this derivation that we do not
yet need to make any assumptions about the behavior of $a(y)$ as
$y\to\pm\infty$ in order to obtain (2.18).

With this second derivation of
(2.18) in hand we can examine the behavior of $f(x,\mu_0^2)$ near the
kinematic limits $x=0$ and $x=1$.  The distribution function $f(x,\mu_0^2)$
must  vanish for $x>1$ since there are
no physical intermediate states in this range.  The natural
constraint $a(1) = b(1) = \ldots 0$ suffices to guarantee that $f(x,\mu_0^2)$
vanishes at $x=1$ independent of $\epsilon$.

The function $f(x,\mu_0^2)$ need not vanish
for $x<0$ -- for negative values of $x$, it is related to
$\bar{f}(-x,\mu_0^2)$.  However, the limit $x\to 0$ corresponds to $y\to
-1/\epsilon$ in (2.19) and this is a physically interesting limit from the
perspective of the heavy quark expansion.
By comparing the arguments of the $\delta$-functions in (2.17) and
(2.2), we
see that $x \to 0$ corresponds to
spectators  (light quarks and glue)
with momentum $P^\mu_\chi \sim mv^\mu$,
since  $P_\chi\cdot n =
\epsilon(1-y)$.  Since specatators with momenta of order $mv^\mu$ are not
allowed within the framework of the heavy quark effective
theory, we cannot study the $x\to 0$ limit in the context of the
heavy quark expansion.  However, we can conclude on the physical basis of
the heavy quark expansion that the functions $a(y)$, $b(y)$, {\it
etc.\/}  vanish rapidly as $y\to -\infty$.  Any more detailed statement
is beyond the scope of the heavy quark effective theory.

In the following section, we will describe a formulation of the heavy quark
expansion in which the parameters $A_k$, $B_k$, {\it
etc.\/} play a central role.  The assumption that these coefficients exist
up to some index $k_{max}$ requires that the functions $a(y)$, $b(y)$, {\it
etc.\/} vanish  faster
than $y^{-k-1}$, $y^{-k}$, {\it etc.\/}, respectively, as $y\to -\infty$.  If
$a(y)$, $b(y)$,  {\it etc.\/} vanish faster than any power of $y$ in this
limit, then of course, the $A_k$, $B_k$, {\it etc.\/} exist for all $k$.  If
they vanish only like some power, then the parameters $A_k$, $B_k$, {\it
etc.\/} exist up to some $k_{max}$.  We cannot determine $k_{max}$ from the
heavy quark expansion; we take it to be a parameter of our analysis on which
we comment when appropriate.

Since all mass dependent factors have been scaled out of $A_k$, $B_k$ {\it
etc.\/}, these should all be of order unity.  In fact, in free field theory
all the $A_k$ would equal unity and $a(y)$ would be a $\delta$-function at
$y=1$, corresponding to $P^\mu_\chi = 0$, in (2.2) ({\it i.e.\/} no
spectators).  In the real world the heavy quark is not free and $a(y)$ is not
a $\delta$-function.  However, since $a(1)=0$ and $a(y)\to 0$ as
$y\to-\infty$, it is clear that (2.18) describes a function, $f(x,\mu_0^2)$,
which
is sharply peaked near $x=m/M\approx 1-\epsilon$ and whose width
is of order
$\epsilon$.  In fact, since $A_0 = 1$, $f(x,\mu_0^2)$ approaches a
$\delta$-function as $\epsilon\to 0$, reproducing free field theory in this
limit as it should.  It is important to appreciate the fact that $a(y)$
provides a heavy-quark-mass independent measure of the influence of
confinement
on the heavy quark momentum distribution in a heavy meson --- a fact which
could provide tests of the heavy quark expansion.
This will be discussed below.

Equation~(2.18) gives $f(x,\mu_0^2)$ at the heavy quark mass scale.
Experiments
will generally be done at a value of momentum squared, $Q^2$, higher than
this.  What we have computed so far is simply the ``boundary data'' for QCD
evolution from a low mass  scale to the scale $Q^2$.  This evolution
is non-trivial and complicates the comparison of (2.18) with experiment.
We postpone the important discussion of comparison with
experiment until we have considered the fragmentation function.
\goodbreak
\bigskip
\noindent{\bf II.2\quad Fragmentation Function}
\medskip
\nobreak
The analysis is very similar to that of the previous section.
We first
define the heavy quark fragmentation function as the leading
twist contribution of the heavy quark to the process
$e^+ e^- \to H(P) + X$ where
$H(P)$ is a stable meson with four-momentum $P_\mu$
containing the heavy quark $h$ and $X$ is any
additional hadronic contribution to the final state.  We first give the
analog of (2.1) for the spin averaged fragmentation function.$^7$
$$\hat f(x,\mu^2) = {1\over 4\pi} \sum_\chi \int
d\lambda\,e^{i\lambda x} \ll 0 \left| n_\alpha \gamma^\alpha_{ij} h_j (\lambda
n)\right|H(P)\chi\rr \ll H(P)\chi|\overline{h}_i(0)|0\rr\ \ \eqno(2.23)$$
where we have explicitly incorporated the Lorentz indices.
As before, we suppress $\mu^2$ on the right-hand side and assume $n\cdot A=0$
gauge.  Here, because we cannot perform the sum on $\chi$, the relevant
product
of matrix elements does not reduce to a bilocal operator.
As with the structure function, the process
$e^+e^-\to H(P) + X$ also receives a contribution from
$\hat{\overline{f}}(x,\mu^2)$ which measures anti-$h$-quark fragmentation
into $H$ plus anything.  This should be negligible
for $\mu^2\sim m^2$.
For now, we retain the variable $x = Q^2/2P\cdot Q$ ($1<x<\infty$), which in
the center of mass frame is the ratio of the beam energy to the energy of the
hadron.  With this variable, $\hat f(x,Q^2)$ can be interpreted as the
probability of finding a hadron of momentum fraction $1/x$ when a quark of
unit momentum
is created from the vacuum by
a current of virtuality $Q^2$. As before, this  can be seen by translating
$h(\lambda n)$ to $h(0)$ and using light cone projection operators,
$$\hat f(x,\mu^2) = {1\over 2p} \sum\limits_\chi\delta\left( x - 1 -
P_\chi\cdot n\right) \left| \ll H(P)\chi\left|
h^\dagger_+(0)\right|0\rr\right|^2 \ \ .\eqno(2.24)$$

The evaluation of $\hat f(x,\mu^2)$ proceeds similarly to that of
$f(x,\mu^2)$.  Here we define
$$\sum\limits_\chi \ll 0\left|n_\alpha\gamma^\alpha (n\cdot\Pi)^k
h_v(0) \right|H(P)\chi\rr \ll H(P) \chi \left|\overline{h}_v(0)\right|0\rr =
2\epsilon^k A_k\eqno(2.25)$$
and similarly for the subleading terms.  We also define
functions $\hat{a}(y)$ and $\hat{b}(y)$ by
$$\hat A_k = \int^\infty_1 dy\,y^k\,\hat a(y)\ \ ,\qquad \hat B_{k+1} =
\int^\infty_1 dy\,y^k\,\hat b(y)\ \ ,\eqno(2.26)$$
{\it etc\/}.

The range of $\hat a(y)$ has been chosen so that the
fragmentation function will only be non-zero in the physical region, $1\le
x <\infty$.  Following an identical analysis to that for the structure
function, we derive $$\hat f(x,\mu_0^2) = {1\over\epsilon} \hat a \left( {x -
{\displaystyle{m\over M}} \over \epsilon}\right) + \hat b\left(
{x-{\displaystyle{m\over M}}\over \epsilon}\right) + \ldots\eqno(2.27)$$
where $\hat b$ is the subleading term defined analogously to the previous
section and $\ldots$ refers to further suppressed functions.
 With the more conventional definition of the fragmentation function variable,
$z = 1/x$, one obtains
$$\hat f({1\over z},\mu_0^2) = {1\over\epsilon}\hat a \left(
{{\displaystyle{1\over z}}
-
{\displaystyle{m\over M}}\over\epsilon}\right) + \hat b \left(
{{\displaystyle{1\over z}} - {\displaystyle{m\over M}}\over \epsilon}\right)
+ \ldots \
\ .\eqno(2.28)$$
Here $z$ is the fraction of the beam energy carried by the hadron.

Equation~(2.28) summarizes the predictions of the heavy quark
expansion for the fragmentation function at the heavy quark mass scale
$\mu_0^2$.  As in the case of the distribution function, the derivation
seems to require the existence of all the parameters $\hat A_k$, $\hat
B_k$, {\it etc.\/}.  However, as before, there is an alternate
derivation which deals directly with the (2.23) without ever expanding in
local operators.  The result is the same, (2.28), without the restriction
that the integrals defined in (2.26) exist.  All of the remarks made in
the previous section concerning the $x\to 0$ or $y\to -\infty$ limit apply
as well to the fragmentation function in the $z\to 0$ or $y\to\infty$
limit.  We shall generally assume that the parameters $\hat A_k$, $\hat B_k$,
 {\it etc.\/} exist for $k$ up to some $k_{max}$.
To fit the parameters, (2.28)
must be evolved
to a scale $Q^2>\mu_0^2$.  We therefore
discuss QCD evolution in the next section.
\goodbreak
\vfill
\eject
\noindent{\bf III.\quad EVOLUTION, LIMITATIONS OF THE HEAVY QUARK EXPANSION
AND
COMPARISON WITH EXPERIMENT}
\medskip
\nobreak
\noindent{\bf III.1 \quad Moment Analysis}
\medskip
\nobreak
We have obtained the general form for the fragmentation function
at a low renormalization scale. Because the fragmentation
function is measured at $Q^2 \gg m^2$, it is necesary to discuss the
QCD evolution between the two scales.  This is
most conveniently accomplished by evolving the
measured moments of the fragmentation
function to the scale $\mu_0^2$
at
which we have parameterized the function.
At this scale, one  extracts the mass independent parameters.

In this subsection,
we will discuss
the application of this procedure to the first few moments
of the fragmentation function.
In
later subsections, we explain why we restrict our
analysis to only these lowest moments. We
parameterize these moments in terms of mass independent
parameters (related to those which entered the functions $\hat a(y)$ and
$\hat b(y)$) and show furthermore
how to extract the parameters charactizing the moments at
leading and subleading order in $\Lambda/m$.

We define the moments of the fragmentation function
at a scale $Q^2$ above the heavy quark scale
in the usual way.
$$\hat\Gamma_k(Q^2) = \int^1_0 dz\, z^k \hat f({1\over z},Q^2)\ \
.\eqno(3.1)$$ Notice we study the moments in $z$ for positive $k$.
These are the  conventional
moments:  they scale simply under QCD evolution and are convergent
at all scales $Q^2$.
Having defined the moments at the scale $Q^2$, one must
renormalize them from $Q^2$ to the scale  $\mu_0^2$, where we have
parameterized
the matrix elements.  This proceeds in three steps.  First,
one must evolve the function from the high momentum scale
$Q^2$ at which the experiment takes place to the heavy quark mass
scale $m^2$.  Then one must match the evolved moments onto
the heavy quark theory. Finally, one must evolve in
the heavy quark theory from the scale $m^2$ to the scale $\mu_0^2$ at
which the matrix elements were parameterized.  To apply heavy
quark flavor symmetry, this scale should be the same for the
$b$ and $c$ quarks.

The first stage of evolution, between $Q^2$ and $m^2$,
is standard. The fragmentation function measured in $e^+ e^-\to H+X$
at the large scale $Q^2$
is a linear combination of quark, antiquark and
gluon fragmentation functions.  The gluon and antiquark fragmentation
functions are not zero at the scale $Q^2$.  However, they evolved from
the scale $m^2$ at which both the antiquark and gluon fragmentation
functions were negligible.  In particular,
because the quark is heavy,
the gluon moments at the scale $m^2$ are suppressed
by $(\Lambda/m)^2$ relative to the quark contribution.
With these constraints on the antiquark and gluon fragmentation
functions it is possible to de-evolve the data from $Q^2$ using the
standard anomalous dimension matrix and obtain the heavy quark
fragmentation function at the scale $m^2$.
Above the heavy quark mass scale, the fact
that the quark is heavy is irrelevant
to the evolution equations, and enters only
in the boundary condition which tells us the
gluon moments are small at the scale $m^2$.
This information about the boundary data, together with the experimental
data at the scale $Q^2$, is  sufficient to solve completely the evolution
equations.

The second stage in the
procedure is to match from the full theory onto the heavy quark theory.  From
the definition of the fragmentation function given in (2.23), it is
clear that the moments match onto the moments in the heavy quark theory
in a straightforward way.
That is, we already know how to match
the matrix elements appearing in the function
itself onto matrix elements of the heavy
quark effective theory characterized by the parameters
$\hat A_k,\hat B_k,\ldots$ which we have defined.
The moments are  defined as integrals of $\hat f(x,\mu_0^2)$ times
powers of $x$. This definition holds in both
the full and effective theories. If
one were to work to subleading order, there would be additional
contributions from evaluating the one loop matrix elements
in the full theory.

The final stage is to de-evolve the moments
in the heavy quark effective theory from the heavy quark mass scale,
$m^2$, to the uniform scale of the effective theory, $\mu_0^2$.
Fortunately, to leading and sub-leading order in $\epsilon$, this
is trivial.  The point is that the matrix elements at leading
and subleading order in the heavy quark mass expansion {\it
do not evolve}.  This is most
easily seen in  $v \cdot A=0$ gauge
where $v_\mu$ is the heavy meson velocity.
Because the parameters  $\hat A_k$
and $\hat B_k$ are defined in terms of matrix elements of local,
gauge--invariant operators in the effective theory which involve only  a
single velocity $v_\mu$, the gluons do not couple in leading order in the heavy
quark mass expansion. Of course, at the level of $1/m^2$, there are QCD
radiative corrections in the heavy quark effective theory.  Therefore, in
order to extract parameters at order $\epsilon^2$ further QCD evolution would
be necessary.

To summarize, the moments of the heavy quark fragmentation function should
be determined experimentally from the data at the scale $Q^2$.  They should
then be de-evolved via QCD evolution down to the scale
$m^2$, at which point they should be matched to the moments of
the function $\hat f(x,m^2)$ parameterized in the heavy quark
effective theory through the parameters $\hat A_k$ and $\hat B_k$,
or equivalently, the functions $\hat a(y)$ and $\hat b(y)$ which
summarize the relations among the moments.
The QCD radiative corrections in the
full theory are given in Ref.~[5] to subleading order.  For the leading
parameters no further QCD evolution in the heavy quark effective theory is
required.

For simplicity, we therefore compare
the evolved, experimentally determined
moments at $\mu^2=m^2$ to
the moments derived from the function (2.23).
It is more convenient
to work with the variable $x$ to which we now revert.  Then
$$\eqalign{\hat\Gamma_k (m^2) &= \int^\infty_1 dx\,x^{-k-2}\,\hat f(x,m^2)
\cr &= \int^\infty_1 dx\,x^{-k-2} \left( {1\over \epsilon}\hat a \left({x -
{\displaystyle{m\over M}}\over\epsilon}\right) + \hat b
\left({x-{\displaystyle{m\over M}}\over\epsilon}\right)+\ldots \right).
\cr}\eqno(3.4)$$
A simple change of variables yields
$$\eqalign{\hat\Gamma_k(m^2) & =\int^{\infty}_1 dy \left(\epsilon y + {m \over
M}\right)^{-k-2}\left(  a(y)+\epsilon b(y)+\ldots \right) \cr &=
\sum\limits^\infty_{\ell=0} \left(\matrix{-k-2\cr \ell\cr}\right) \left\{
\epsilon^\ell \tilde A_\ell + \epsilon^{\ell+1} \tilde B_{\ell+1} +
\ldots\right\} \cr}\eqno(3.5)$$
where we have shown explicitly  only those functions which
begin at order $\epsilon^0$ or order $\epsilon$.  Here
$$\eqalign{\tilde A_\ell &\equiv \int^\infty_1 dy(y-1)^\ell a(y) \cr
\tilde B_{\ell+1} &\equiv \int^\infty_1 dy(y-1)^\ell b(y)\ .\cr}\eqno(3.6)$$
Notice that only a finite
number of $\tilde A_\ell$, $\tilde B_\ell$,
{\it etc.\/} occur in the expansion of any
given moment to a fixed order in $\epsilon$.

Consider now the explicit expressions for the first few moments.
$$\eqalign{
\hat\Gamma_0(m^2)
&= \tilde A_0 +\epsilon\left(-2 \tilde A_1+\tilde B_1) + \epsilon^2 (3
\tilde A_2 -2 \tilde B_2 +\tilde C_2\right) + \ldots
\cr
\hat\Gamma_1(m^2) &= \tilde A_0 + \epsilon
\left(-3\tilde A_1 + \tilde B_1\right) +
\epsilon^2\left(6\tilde A_2-3 \tilde B_2 + \tilde C_2\right) + \ldots \cr
\hat\Gamma_2(m^2) &= \tilde A_0 + \epsilon\left( -4\tilde A_1 + \tilde
B_1\right
) +
\epsilon^2\left(10\tilde A_2
-4\tilde B_2 + \tilde C_2\right) + \ldots\cr
\hat\Gamma_3 (m^2)
&= \tilde A_0 + \epsilon \left( -5\tilde A_1 + \tilde B_1\right) +
\epsilon^2\left(15\tilde A_2 -5\tilde B_2 + \tilde C_2\right) + \ldots
\cr}\eqno(3.7)$$
If we work to order $\epsilon$ and require the remainders in (3.7) to be
bounded, then it is necessary to assume that $a(y)$ vanishes faster
than $y^{-3}$ for large $y$.  Likewise $b(y)$ must vanish faster than
$y^{-2}$.  In order to work to order $\epsilon^2$, $a(y)$ and $b(y)$ must
vanish one power faster for large $y$.
Assuming these asymptotic forms, and that  the heavy quark expansion is
valid, we can extract the leading parameters of the functions $\hat a(y)$
and $\hat b(y)$.  In particular, from knowledge of either the $b$ or $c$
quark fragmentation function alone, one can extract the parameters $\tilde
A_0+\epsilon \tilde B_1 + \epsilon^2 \tilde C_2+\ldots\approx \tilde A_0$,
$\epsilon \tilde A_1-\epsilon^2 \tilde B_2
+\ldots\approx \epsilon\tilde A_1$, and $\epsilon^2 \tilde A_2$,
working only to order $\epsilon^2$.  With knowledge of both the $c$ and $b$
quark fragmentation functions, one can independently extract
$\tilde A_0$, $\epsilon \tilde B_1$, $\epsilon \tilde A_1$, $\epsilon^2 \tilde
B_2$, and $\epsilon^2 \tilde A_2$.
Of course if we work only at order $\epsilon$, there is an
equal spacing rule which serves as a consistency check on the moments:
$$\hat\Gamma_1 - \hat\Gamma_0 = \hat\Gamma_2-\hat\Gamma_1=\hat\Gamma_3-
\hat\Gamma_2\ \ .\eqno(3.8)$$
\medskip

We can now also see the main problem with  extracting
information from higher moments.  In order to extract
a  parameter at any specified order in the heavy quark mass
expansion, we need to assume that higher order terms can be neglected.
However, the large binomial coefficients multiplying higher order
terms in $\epsilon$ make this impossible for the higher moments.
That is, for the $k$th moment, with $k$ large, the expansion
is not in $\epsilon$, but in $k\epsilon$, which compromises
the application of the expansion
beyond the lowest moments. In the next sub--section,
we discuss the further advantages to be had by restricting
attention to only the lowest moments.  After that (in III.4), we return to
the question of whether information can be extracted from the higher moments,
or equivalently from the function $\hat f(x,m^2)$ itself.

\goodbreak

\noindent{\bf III.2 \quad Subleading QCD Evolution }\medskip
\nobreak

We have discussed leading logarithmic QCD evolution.
However, it is well known that the usual QCD evolution
program has difficulties with heavy
quark fragmentation functions. In this section,
we explain why these problems are less severe for our program than for a
fully perturbative study of heavy quark fragmentation functions and may be
ignored entirely if we restrict our analysis to the lowest moments.

In principle, one would expect that QCD evolution in the
full theory between $Q^2$ and $m^2$ is perturbative,
as both scales are much higher than the QCD scale.  However,
near $x=1$, the QCD coupling $\alpha_s$ is multiplied
by logarithms of the form $\log(1-x)$, etc.   In
this region, effects which are formally subleading are
nevertheless large.  In fact, it has been shown that the
summation of these Sudakov logarithms
can be summarized by evaluating $\alpha_s$ at the scale $(1-x)Q^2$.$^8$
  These effects are important when $(1-x)Q^2
<\!\!< Q^2$ and furthermore the analysis suggests that other, truly
non-perturbative effects cannot be neglected for $x$ such that
$(1-x)Q^2\sim\Lambda^2$.

Now consider the form of the function renormalized at $m^2$ given
by (2.28). It
 vanishes at $x=1$ and is peaked at $x-1\sim \epsilon$.  Sudakov effects
are worst at the lowest $Q^2$, in this case $Q^2=m^2$.
At $Q^2=m^2$,
non-perturbative effects beyond the resummation of Sudakov logarithms are
important for $1-x<\epsilon^2$.  The distinction between $\epsilon$ and
$\epsilon^2$ is crucial. Since $\hat
f$ is likely to be a rapidly falling
function in the region close to $x=1$, and since the scale for variation in
$\hat f$ is $\epsilon$, the function should be very small in the region
where Sudakov logarithms are important. In particular,
the first few moments
are very likely dominated by $x$ in the regime where even Sudakov logarithms
are
not large, so a leading logarithmic analysis should be valid. However, we
do not know how fast $\hat f$ falls as $x\to 1$. This non-perturbative
information determines which moments are trustworthy.  Since
constraints on the heavy quark expansion
most likely restricts us more
severely to low moments than do considerations of Sudakov effects, we
expect that Sudakov logs can be neglected altogether.

This is in contrast to
previous works$^{4,5}$ which attempted
a fully perturbative analysis of the
fragmentation function over the whole range of $x$.
They have been led to a form
of the fragmentation function which is singular at $x=1$ and have necessarily
had to deal with evolution in a regime in which non-perturbative effects are
important.  From this perspective, our approach has been to summarize the
nonperturbative information in a few parameters.  With sufficiently rough
averaging (by looking only at
low moments) we can then do a simpler QCD analysis.

We see now the essential advantage gained by restricting attention
only to the lowest moments, independent of the breakdown of the $\epsilon$
expansion.  Our objective is to cleanly distinguish perturbative
and non-perturbative QCD.  For $x$ very close to 1, this distinction
is not possible.  But this region
of $x$ should not contribute significantly to the low moments
for the reasons outlined above.

Having addressed the issue of Sudakov logarithms, the
study of higher order QCD effects should be standard.  For
example, there is an ambiguity associated with the precise
scale at which we match between the full and heavy quark
effective theories which can only be resolved through a higher
order QCD calculation. Moreover, as a matter of practice,
the data on the $c$ quark is obtained at center of mass
energy of the $\Upsilon(4S)$  Furthermore, it might be that in this case the
logarithm is not sufficiently large for a leading log analysis to be adequate.
Certainly, a subleading calculation would be required to extract a parameter at
the precision of $\epsilon^2$. \medskip
\noindent{\bf III.4 \quad Direct Analysis of $\hat f(x,m^2)$}
\medskip

Having analysed the limitations on the moment analysis, we return briefly to
the possibility of extracting information from looking directly at the
function $\hat f(x,m^2)$.  As long as we restrict $(1-x)>\!\!>\epsilon^2$
Sudakov effects are not likely to be important.  Looking at (3.7) we see that
the moment analysis breaks down when $k\epsilon>1$.  However, the terms which
become large in this limit can be resummed yielding the original functional
form, (2.28), in which successive terms are suppressed by powers of $\epsilon$
when the argument, $y=({1\over z}-{m\over M})/\epsilon$, is held fixed.  So it
is appropriate to analyze $\hat f(x,m^2)$ directly.

First, suppose the function $\hat f(x,m^2)$ can be extracted from experiment
and QCD de-evolution. The predictive power of (2.28) lies in the orderly
expansion in $\epsilon$ implied for $\hat f(x,m^2)$ when
$x=1+\epsilon(y-1)$,
$$\hat f \left( 1 + \epsilon(y-1),m^2\right) = {1\over\epsilon} \hat a(y)
+ \hat b(y) + \ldots
\eqno(3.9)$$
where $\hat a(y)$, $\hat b(y)$, {\it etc.\/} are all independent of $\epsilon$.
In principle,
$\hat f$ can be measured for both $c$- and $b$-quarks, $\hat a(y)$,
$\hat b(y)$ and
$\epsilon$ can all be extracted and the heavy quark expansion can be tested.
To
leading order in $\epsilon$, (2.28) requires that there is a choice of
$\bar\Lambda$ (which in principle could have been extracted elsewhere$^{10}$)
such that
 $$\epsilon_c \hat f_c \left( 1 + \epsilon_c (y-1),m^2_c\right) -
\epsilon_b\hat f_b\left( 1 + \epsilon_b(y-1), m^2_b\right) = 0
\eqno(3.10)$$
 Having fixed
$\epsilon_c$ and $\epsilon_b$, $\hat a(y)$ can be extracted from
$$\eqalignno{
\hat a(y) &= {\epsilon_c\epsilon_b\over \epsilon_c-\epsilon_b} \left( \hat f_b
\left( 1 + \epsilon_b (1-y), m^2_b\right) - \hat f_c \left( 1 + \epsilon_c
(1-y), m^2_c\right)\right) &(3.11)\cr}$$
A cautionary note is appropriate: After de-evolution
$\hat f(x,m^2)$ would likely be available only with some finite
resolution (smearing) considerably greater than the original experimental
resolution at $Q^2$.  If the resolution were as large as the intrinsic width
of $\hat f(x,m^2)$, which is of order $\epsilon$, then it is easy to see that
this method loses its usefulness and the moment method outlined previously
gives
as much information as can be extracted from the data.
\medskip
\noindent{\bf III.3 \quad Discussion}
\medskip
\nobreak
There are several points we have neglected in the previous analysis
which we now clarify.  First, we have made assumptions about the
behavior of the heavy quark and antiquark distributions at the heavy
quark mass scale. In particular, we assumed $\hat f(x,m^2)$ vanishes
rapidly
as $x\to\infty$ and we have ignored $\hat{\overline{f}}(x,m^2)$.
These assumptions are only sensible if they are stable
under perturbative QCD evolution in the heavy quark theory.
They would clearly not be consistent above the heavy quark scale,
 at which point conventional evolution
generates a tail in $\hat f$  which reaches to $z=0$, a gluon
distribution at order $\alpha_s\ln(Q^2/m^2)$, and a heavy antiquark
distribution $\hat{\overline{f}}$ at order $\alpha_s^2\ln^2(Q^2/m^2)$.
Thus it would not be reasonable to implement such assumptions in the
domain of perturbative evolution.  However, we are placing these
restrictions on the {\it boundary data\/} at and below the heavy
quark mass scale, $m^2$.
As we have already discussed, the structure function does not evolve
at scales below $m^2$ at leading or subleading order in the heavy
quark mass expansion.  At this level of accuracy, our
assumptions are therefore stable with respect to with perturbative QCD
evolution.

We have also neglected gluon and antiquark contributions
to the moments at the scale $m^2$. Because a gauge invariant
gluon operator matrix element is suppressed by at
least $\epsilon^2$ and receives no enhancement
by a binomial coefficient relative to the leading
and subleading terms, it can only contribute to
the function $c(y)$ and other suppressed functions.
At the level in the heavy quark expansion at which we
are working, we can neglect the gluon moments
at the scale $m^2$.  However, as we have
discussed, gluon moments are generated through
perturbative QCD evolution so they do not vanish
at the scale $Q^2$.  Similarly, antiquark moments
are only generated through QCD evolution.

The next point to clarify is that throughout our analysis, we
have assumed fragmentation into
a specific final state, namely the pseudoscalar meson.
However, experiments at very high center of mass energy will
not distinguish the pseudoscalar and vector $B$ mesons.
Furthermore, the final state could be a $B_s$
meson or a baryon containing a $b$ quark.
However, the fragmentation function into $D^*$
is what is best measured at CESR.

If we are to apply our methods to compare the fragmentation
function at
LEP and CESR,  we would not expect to obtain results
which are accurate to better than order 20\%.  This is because heavy quark
symmetry only guarantees the identity of the matrix
elements between the pseudoscalar or the vector states at zeroeth order in
$1/m$.
This means that the parameter $\epsilon\tilde B_1$ is not identical for
the pseudoscalar and vector mesons.  However, counting
spins shows that the vector states are produced three times
as often as the pseudoscalar.  So even if the time ordered
product of the spin splitting operator and the zeroth order
matrix element accounted for 50\% of the contribution to $\epsilon\tilde B_1$,
the relations between the $B$ system and the $D$ system (namely
the vector states) should be accurate at the 15 to 20 \% level,
which is the inaccuracy represented by higher order terms in the
heavy quark expansion.
(The mass splitting between
the pseudoscalar and meson masses is higher order, and can be safely
neglected.)

$B_s$ production should also
not be a problem at this order of accuracy. The $B_s$ is only
produced about 12 \% of the time, and chiral symmetry should
guarantee that the matrix elements are equal to those of the $B_d$
at the 20\% level.

Finally, baryons will probably be produced
about 10\% of the time.
This can be a problem since the parameters of the baryons
are independent of those of the mesons.
The danger is that an inaccurate
determination of $\tilde A_0$ would lead to a large error in extracting
 $\epsilon \tilde A_1$ and $\epsilon \tilde B_1$.
However, there is only a single new parameter at the level
of accuracy of our calculation, namely the fraction of baryons, $p$
(since $p \epsilon$ is already only a few percent.).
If a guess of $p$ at about 10\% is accurate
at the few percent level, one should still be able
to extract the same parameters as before. If $p$ is
left a completely free parameter, $\epsilon \tilde B_1$ can still be extracted.
And of course if the baryons can be excluded from the
sample, everything would proceed as in the text. In fact,
the same analysis would apply to baryons as to mesons.

Furthermore, if it is possible to study
$D$ mesons without restricting oneself
to $D^*$'s, but instead sums inclusively
over all states,  the relative fraction of mesons
and baryons should be independent of heavy quark mass
 at leading order
in the heavy quark mass expansion, so that
the parameter $p$ need not be extracted. The analysis
we gave would apply, but with each parameter in the moment
expansion interpreted as a sum of a meson and baryon contribution.

It should be kept in mind that the consistency tests
among the moments of any particular flavor quark should
hold even with different final state particles, since
they only required the existence of a heavy quark expansion,
which is true for each final states individually. It is
only when comparing between different heavy quark systems
with different fractions of the various final states
that complications arise, unless one can identify the
contributions from the particular final states.

Notice also  that we have expanded in $\epsilon$, which
scales inversely to the meson mass so the expansion is not
strictly in heavy quark mass. It is straightforward to convert
our results to an expansion in heavy quark mass.

Of course, there are potential experimental difficulties
which we expect could make an analysis beyond the level
of $\epsilon$ impracticable. In particular, finite experimental
resolution could be a problem. The spread of the measured
fragmentation function of the $b$ is increased
significantly by the perturbative QCD
contribution. The measurement has to be sufficiently accurate
to determine the nonperturbative parameters once the perturbative
contribution has been removed. Furthermore, what is measured
is the hard lepton in the final state.  With no further theoretical input one
needs an accurate relationship between the momentum of this lepton and the
momentum of the meson itself in order
to accurately measure the fragmentation function.
This requires good measurements of the decay lepton spectrum
from $B$ decay. Alternatively, one would require
a large sample of fully reconstructed events.  However with the heavy quark
expansion one could determine the lepton energy distribution, thereby
simplifying the analysis.

Finally, we comment on the commonly used Peterson fragmentation
function. It is clear from any analysis of heavy quark
fragmentation functions that one expects a function
which is weighted towards 1 but with some spread, and which
goes to zero at $z=0$ and $z=1$.  Any function of this
sort will more or less fit the data. So the Peterson function
is as good as any other. However, the Peterson function
assigns a width of order $\epsilon^2$, where we have
seen the true width of the fragmentation function is order $\epsilon$.
Therefore, one must be careful in assigning a physical interpretation
to the parameters of the fragmentation function, or in trying
to relate the fragmentation functions for different flavors of
quark.

\goodbreak
\medskip
\vfill
\eject

\noindent{\bf IV.\quad CALCULABLE HIGHER-TWIST CORRECTIONS}
\medskip
\nobreak
We now turn to the subject of higher-twist corrections to $e^+e^-\to
H(P)+X$. In general, the study of higher-twist effects is theoretically
complicated and experimentally difficult.  In this case, however, the heavy
quark mass sets the scale for higher-twist effects which are
potentially the most important
because they are of order $m^2/Q^2$ which
in certain experiments can be large ({\it
e.g.\/} $m_b\cong 5$~GeV, $Q\cong 15$~GeV gives $m^2/Q^2\rapp 10\%$).
Fortunately
they can be related to the leading twist fragmentation function.
Because the subject is technically complicated, we first summarize our results
for
the reader who does not wish to see the derivations.  We next give a simple,
non-technical derivation of the primary results.  Finally we summarize the
results
of a complete, operator-product expansion through twist-four $\left(
{\cal O} (1/Q^2)\right)$ distribution
functions which will be published elsewhere.$^{10}$

First, our conclusions:
At leading order in $\alpha_s$, we
 show that there are no power corrections to
$e^+e^-\to H(P)+X$  which scale by the heavy quark mass alone, that is, no
corrections of the form $\left( m^2/Q^2\right)^k$ for $k\ge1$ to either the
transverse or longitudinal fragmentation functions defined by
$$\eqalignno{ \hat W_{\mu\nu} &= {1\over 4\pi} \sum_X \int d^4\xi\,
e^{iq\cdot\xi}\ll 0 \left| J_\mu (\xi) \right| H(P)X\rr\ll
H(P)X\left|J_\nu(0)\right|0\rr &(4.1) \cr
&= - \left( g_{\mu\nu} - {q_\mu q_\nu\over q^2}\right) \hat{w}_L \left(
q^2,\nu\right) \cr &+ \left( g_{\,u\nu} - {P_\mu q_\nu + P_\nu q_\mu\over \nu}
+
 {P_\mu P_\nu q^2\over \nu^2} \right)\hat w_T \left( q^2,\nu\right), \
\ &(4.2) \cr}$$ where $\nu\equiv P\cdot q$.
[The ``fragmentation function'' discussed in Section~II is the scaling limit
of $\hat w_T(q^2,\nu)$.  The function
$\hat w_L(q^2,v)$ vanishes at leading twist
on account of a Callan--Gross-like relation.]
At higher order in $\alpha_s(Q^2)$, there are radiative corrections
which yield
 twist-four corrections of the form
$${m^2\over Q^2}\left( {\alpha_s(Q^2)\over \pi}\ln {Q^2\over m^2}\right)\ \
$$ (as well as nonvanishing $\hat{w}_L(Q^2)$).
Because of heavy quark constraints on matrix elements at the heavy quark scale
it is possible to calculate the magnitude of these corrections exactly in
terms of the leading twist heavy quark fragmentation function described in
Section~II.  There are also corrections of order $m\Lambda/Q^2$ which can
also be related to the leading twist fragmentation
function and are calculable if $\Lambda$ is known.

We conclude that the process $e^+e^-\to H(P)+X$ will probably
not have large higher twist corrections, even at relatively
low center of mass energy.
  At
heavy meson threshold, $Q^2=4M^2$, ordinary higher-twist corrections are
suppressed by $\Lambda^2/Q^2 \approx \Lambda^2/4M^2<\!\!<1$ so
these are cleary small.
Corrections of order $m\Lambda/4M^2$ are only of order 10 \% for
the $c$ quark and smaller still for the $b$ quark. However, this could
confuse the extraction of an order $\epsilon^2$ parameter.  Corrections of
order $m^2/4M^2 \left( \alpha(4M^2)/\pi\right)\ln\left( 4M^2/m^2\right)$ are
less than 5\% for the $c$-quark and much smaller for the $b$-quark. These
should
be well under control.

Our results follow from simple considerations of the structure of heavy quark
mass corrections to $e^+e^-\to H(P)X$.  The analysis is conceptually simpler
for the deep inelastic scattering process $eH(p)\to eX$ where one can use the
language of the operator product expansion.  We will present our argument for
deep inelastic scattering but it
should hold equally well for  $e^+e^-\to H(P)X$.

We first show that there are no corrections of the form $(m^2/Q^2)^k$.
We work to lowest order in $\alpha_s$.  Since we are interested in effects of
order $\left(m^2/Q^2\right)^k$, we may also set $\Lambda=0$.
 Thus, we may
ignore QCD entirely and calculate $eH(P)\to eX$ from a free, on-shell heavy
quark $h$ as shown in Fig.~1.
This calculation, though trivial, gives us the information
we require. This is true only because the relevant matrix elements
are equal to their free field values up to non-perturbative
corrections suppressed by $\epsilon$.  So at order $\epsilon^0$
and $\alpha_s^0$, a free field theory analysis suffices.

Direct calculation of Fig.~1 gives
$$\eqalignno{\hat W_{\mu\nu} &= {1\over 2}\sum_{\rm spins} \bar{u} (P)
\gamma_\mu \left(\thru q + \thru P + m\right) \delta\left[ \left( q +
P\right)^2 - m^2\right] \gamma_\nu u (P) &(4.5) \cr
&= {1\over 4}\Tr \left(\thru P + m\right) \left( \gamma_\mu \thru q +
2P_\mu\right) \delta\left( q^2 + 2q\cdot P\right)\gamma_\nu \cr
&= {1\over 2}\left( - g_{\mu\nu} + {P_\mu q_\nu + P_\nu q_\mu\over\nu}-
{P_\mu P_\nu q^2\over \nu^2}\right) \delta (x-1) &(4.4) \cr}$$
which is manifestly independent of the quark mass.  Notice
that because we have done this
calculation in free field theory,
 the structure function itself is  trivial: $f_1(x) =
{1\over 2} \delta(x-1)$ in the limit $\alpha_s$, $\Lambda\to 0$.  This is the
structure function of a free quark, the inverse Mellin transform of the matrix
elements of the only twist-two operators which persist in this extreme
limit:
$${\cal O}^n_{\mu_1\ldots \mu_n} = {\cal S}_n \overline{h} \gamma_{\mu_1}
\Pi_{\mu_2} \ldots \Pi_{\mu_n} h - \hbox{traces}\ \ \eqno(4.5) ,$$
all of which have identical coefficients ($A_k=2$) in the operator product
expansion.
The matrix elements of the ${\cal O}^n$ will, in reality, deviate from the
free field limit by corrections of order $\Lambda/m$ --- already described in
Section~II --- but these modifications will not alter the relations among
coefficient functions which result in the cancellation of all ${\cal O}\left(
m^2/Q^2\right)^k$ terms from (4.4).

To understand this result more deeply and to see why the corrections of order
\hfil\break
${m^2\over Q^2}\left( {\alpha_s(Q^2)\over\pi} \ln {Q^2\over m^2}\right)$
and $\Lambda m/Q^2$ are calculable, it is necessary to consider the process
from the viewpoint of the operator product expansion.  There are many sources
of $Q^2$ suppressed
 corrections.  The situation is made more tractable by
working at tree level, postponing the study of the effects of evolution and
operator mixing which will of course generate corrections of order
$\alpha_s$ to this result.
  At tree level and through order $1/Q^2$
there has been a complete analysis of electroproduction with {\it massless\/}
quarks.$^{11}$  The twist-four operators which appear in that
analysis include four-quark operators, generically $(\overline{h}h)^2$, two
quark gluon, $\overline{h}Fh$, and two quark two gluon, $\overline{h}FFh$,
operators.  It is easy to see that, renormalized at a scale
where the heavy quark effective theory applies,
these operators contribute only at order $\Lambda^2/Q^2$,  The dangerous
terms are those operators directly proportional to $m$ or $m^2$ which were
omitted from the calculation of Ref.~[11].  Four categories of such
contributions occur:
\medskip
\item{1.}``Target mass corrections'' --- These well-known kinematic
corrections were first understood completely by Nachtmann.$^{12}$
They arise when twist-two operators --- traceless and symmetric in all
indices --- are projected out of the dominant operators which appear
naturally in the expansion of two currents.  Target mass corrections are
proportional to $\left( M^2/Q^2\right)^k$ where $M$ is the heavy meson mass,
and their $x$-dependence is determined by the leading twist-two distribution
or fragmentation function.
\medskip
\item{2.}Modifications of the struck quark propagator --- The replacement
$1/p^2 \to 1/(p^2-m^2)$ generates corrections of order $\left(
m^2/Q^2\right)^k$ to the coefficient functions of the leading twist-two
operators.
\medskip
\item{3.}New twist-three operators with coefficient functions linear in $m$
--- The quark mass term in the Dirac propagator, $(\thru
p+m)/(p^2-m^2)$, generates a tower of twist-three, chiral odd operators of
the form $\bar{\psi}D_{\mu_1}\ldots D_{\mu_n} \psi$ with coefficient
functions linear in $m$.   The matrix elements of these operators are
proportional to $M$ and the resulting corrections to the distribution and
fragmentation functions are of order $mM/Q^2$.
\medskip
\item{4.}When twist-four operators are resolved into a canonical form in
which gluon fields are manifest, it is necessary to use the QCD equations
of motion.  The use of the massive Dirac equation $(\thru D - m)h=0$
generates new twist-four operators proportional to $m$ and $m^2$.

\noindent A careful study of the operator product expansion at tree level
through twist-four shows these are the only corrections through order $m^2$.

 The result of the operator product expansion calculation
for the distribution function is most conveniently expressed as moments of
the transverse and longitudinal structure functions, $F_T$ and $F_L$,
defined by $$\eqalign{ F_T(x,Q^2) &= {1\over 2x} F_2 (x,Q^2) \cr
F_L (x,Q^2) &= {1\over 2x} \left( 1 + {4M^2x^2\over Q^2}\right) F_2 (x,Q^2) -
F_1(x,Q^2) \ \ .\cr}\eqno(4.6)$$
Then our results are
$$\eqalignno{{\cal M}^T_k(m^2) &= \int^1_0 dx\,x^{k-1} F_T (x,m^2) \cr
&= \Gamma_k(m^2) + {1\over Q^2} {k-2\over k}\biggl\{ (k-3) M^2 \Gamma_k(m^2)
- {(k-1)(k+6)\over k} m^2\Gamma_{k-2}(m^2)\cr
&\qquad  + {4k-3\over k}2m M\Lambda_{k-2}(m^2)
\biggr\} &(4.7) \cr\noalign{\vskip 0.2cm}
{\cal M}^L_k(m^2) &= \int^1_0 dx\,x^{k-1} F_L(x,Q^2) \cr
&= {4\over Q^2}\ {k-1\over k} \left\{ - M^2 \Gamma_k (m^2) - {k-2\over
k}m^2 \Gamma_{k-2}(m^2) +\right.\cr &\left. {k-1\over k}2mM\Lambda_{k-2} (m^2)
\right\} &(4.8) \cr}$$
where $k=2,4,6,\ldots$ and $\Gamma_k(m^2)$ and $\Lambda_k(m^2)$ are the matrix
elements of twist-two and twist-three heavy quark operators, respectively (at
the renormalization scale $m^2$):
$$\eqalignno{
\Gamma_k(m^2) &= \ll H(P) \left|\overline{h}\thru n \left(\Pi\cdot
n\right)^{k-1} h \right| H(P)\rr\bigg|_{m^2} &(4.9) \cr
M\Lambda_k(m^2) &= \ll H(P)\left|\overline{h} \left( \Pi\cdot
n\right)^kh\right|H(P)\rr\bigg|_{m^2}\ \ .&(4.10)\cr}$$
The first term in (4.7) is the standard twist-two result.  The other terms in
(4.7) and (4.8) each consist of three parts. The first, proportional to
$M^2\Gamma_k$, are target mass corrections. The second, proportional to
$m^2\Gamma_{k-2}$, contain mass corrections to the struck quark
propagator and operators generated by use of the Dirac equation.  The
third, proportional to $mM\Lambda_{k-2}$, originate in chiral odd,
twist-three operators  and also from use of the Dirac equation.  The
fact that these corrections can be grouped into transverse and longitudinal
contributions is a consequence of the gauge invariance of virtual photon
$H$-meson Compton scattering and is a highly non-trivial check on the
algebra. So is the fact that {\it (4.7) and (4.8) reduce to the free quark
result, (4.4), when the $\left\{ \Gamma_k\right\}$ and $\left\{
\Lambda_k\right\}$ are set equal to their free quark values:\/}
$\Gamma_k(\hbox{free})= \Lambda_k(\hbox{free})=2$.

In general, the $\left\{\Gamma_k\right\}$ and $\left\{\Lambda_k\right\}$ can
be
related by the QCD equations of motion:
$$m^2\overline{h}\thru n (n\cdot\Pi)^{k-1} h = m\overline{h} (n\cdot\Pi)^k
h
 +
m\sum^{k-2}_{\ell=0} \overline{h} (n\cdot\Pi)^\ell {\sigma^\alpha}_\beta
F^{\gamma\beta} n_\alpha n_\gamma (n\cdot\Pi)^{k-\ell-2}h \eqno(4.11)$$
for $k>1$, and
$$m^2\overline{h}\thru n h = m \overline{h} n\cdot \Pi h$$
for $k=1$.  In general ({\it e.g.\/} for light quarks) the third operator in
(4.11) obstructs any attempt to relate $\Gamma_k$ to $\Lambda_k$.  However, at
the heavy quark scale it is easy to see that the matrix element of the third
operator is of order $\Lambda^2/m$.  Define
$$C^{(k)} \equiv \ll H(P) \left|\sum^{k-2}_{\ell=0} \overline{h}
(n\cdot\Pi)^\ell {\sigma^\alpha}_\beta F^{\gamma\beta} n_\alpha n_\gamma
(n\cdot\Pi)^{k-\ell-2} h\right| H(P)\rr\ \ .\eqno(4.12)$$
At the heavy quark scale, numerator factors of $m$ arise only from the
substitution $\Pi^\mu=nv^\mu + k^\mu$, plus one additional factor from the
norm of the Dirac fields.  Each factor of $n_\mu$ contributes a factor $1/M$.
The result is $C^{(k)}\propto m^{k-1}/M^k$, but on dimensional grounds
$C^{(k)}$ scales like mass, so we conclude
$$C^{(k)} = \overline{C}^{(k)} {\Lambda^2 m^{k-1}\over
M^k}\eqno(4.13)$$
to leading order in the heavy quark mass expansion. With this result in hand,
we take the matrix element of (4.11) in a heavy meson state and find
$$m^2\Gamma_k (m^2) = mM \Lambda_k (m^2) + {\cal O}(\Lambda^2) \eqno(4.14)$$
where we have restored the renormalization scale $(m^2)$ as a label on the
operator matrix elements to remind ourselves that (4.14) holds at the heavy
quark scale.  Substituting (4.14) into (4.7) and (4.8) we obtain the dominant
twist-four corrections to the longitudinal and transverse moments at the heavy
quark scale, $m^2$:
$$\eqalignno{
{\cal M}^T_k(m^2) \bigg|_{\rm twist-4} &= {1\over Q^2}\  {(k-2)(k-3)\over
k}\left[ M^2\Gamma_k(m^2) - m^2\Gamma_{k-2} (m^2)\right] + {\cal
O}(\Lambda^2/Q^2) &(4.15) \cr {\cal M}^L_k (m^2) \bigg|_{\rm twist-4} &= -
{4\over Q^2} \  {k-1\over k}\left[ M^2\Gamma_k (m^2) - m^2 \Gamma_{k-2}
(m^2)\right] + {\cal
O}(\Lambda^2/Q^2)&(4.16) \cr}$$ $\Gamma_k(m^2)$ can be expressed in terms of
$A_0$, $A_1$ and $B_1$ (to order $\epsilon$) by means of (3.7).  $m$ and
$M$ are related by $m =  (1-\epsilon)M$.
Altogether
$$M^2 \Gamma_k(m^2) - m^2 \Gamma_{k-2}(m^2) = 2\epsilon A_1 + {\cal
O}(\epsilon^2) \eqno(4.17)$$
so we obtain
$$\eqalignno{
{\cal M}^T_k(m^2)\bigg|_{\rm twist-4} &= 4\epsilon {M^2\over Q^2} A_1
{(k-2)(k-3)\over k} &(4.18) \cr
{\cal M}^L_k (m^2) \bigg|_{\rm twist-4} &= - 16\epsilon {M^2\over Q^2} A_1
{k-1\over k}\ \ . &(4.19)\cr}$$
Equations~(4.18) and (4.19) hold at the heavy quark scale, $Q^2 = m^2$, and at
higher scales, $Q^2>m^2$, modulo QCD radiative corrections which are
discussed below.  Like other results which make use of the mass expansion,
(3.7), they are valid only for low moments, for which $k<\!\!<1/\epsilon$
because
we ignored $k^2\epsilon^2$ compared to $k\epsilon$ in the derivation.

Equations~(4.18) and (4.19) summarize our assertion that the leading
higher-twist corrections to the scaling of distribution functions is of order
$\Lambda M/Q^2$, and is calculable in terms of the parameter
$\epsilon\tilde A_1$, which can, in principle, be extracted from the leading
twist heavy quark distribution function.  From the operator-product-expansion
point-of-view, the cancellation of terms of order $m^2/Q^2$ among their several
sources seems fortuitous. The simple graphical argument given at the beginning
of this section gives a firm physical basis for the cancellation.  It also
assures us that the same result holds for fragmentation functions for which
operator product expansion methods are less straightforward.

Finally, it is necessary to comment on the effect of QCD radiative corrections
on these results.  We focus on the moments, ${\cal M}^T_k(Q^2)$ and ${\cal
M}^L_k (Q^2)$, defined by (4.7) and (4.8).   The operators parameterized by
$\Gamma_k(Q^2)$ and $\Lambda_k(Q^2)$ have different anomalous dimensions so
the cancellation of $m^2/Q^2$ terms, which was complete at $m^2$, is
incomplete at $Q^2$.  The terms which survive are clearly
${\cal O}\left( {m^2\over Q^2}\  {\alpha_s(Q^2)\over\pi}\ln {Q^2\over
m^2}\right).$
To establish the way to compute these terms it is necessary first to examine
the $Q^2$ values of interest.  For $Q^2>\!\!>m^2$ one could use standard,
leading order renormalization group evolution.  However, for $Q^2>\!\!>m^2$
twist-four effects are unlikely to be measurable in the first place.  We are
more likely to be interested in $Q^2\rapp m^2$ where twist-four effects might
be measured.  At such scales there is no reason to ignore ${\cal O}\left(
\alpha_s(Q^2)/\pi\right)$ corrections to coefficient functions.  Fortunately,
there is a direct way to compute all relevant corrections of order
${m^2\over Q^2} \  {\alpha\over\pi}\ln {Q^2\over m^2}\colon$
since $\Lambda$ may be taken to zero in these calculations, they may be
extracted from the Feynman diagram calculation of the appropriate process
involving free, on-shell heavy quarks --- the same method we used to show
corrections of order $m^2/Q^2$ vanish at the beginning of this section.  The
relevant diagrams are shown in Fig.~2.  The processes shown in Figs.~2b -- 2d,
which involve mixing of gluonic and quark-gluon operators in the language of
the operator product expansion, are easily seen to be suppressed by
$\Lambda^2/Q^2$.  Thus it is a matter of straightforward calculation to
compute the set of all important $1/Q^2$ corrections to heavy quark
fragmentation or distribution functions: corrections of order $\Lambda M/Q^2$
have been calculated in this paper; corrections of order
${m^2\over Q^2}\  {\alpha_s\over \pi}\ln {Q^2\over M^2}$
can be obtained from calculation of Feynman diagrams involving on-shell heavy
quarks.

\goodbreak
\medskip
\noindent{\bf V.\quad CONCLUSION}
\medskip
\nobreak

There are only very few places where one can hope to extract mass suppressed
contributions to matrix elements.  The existence of these additional
parameters should be very important from the point of view of testing heavy
quark symmetry at subleading order. Furthermore, it will be useful to see what
the size of mass suppressed corrections really turns out to be
in order  to test
the validity of the heavy quark expansion.

Clearly the ideas of this paper need to be tested by application
to existent and future data.  Future work would involve incorporating
subleading evolution and calculation of the perturbative higher
twist effects.

\goodbreak
\medskip
\noindent{\bf ACKNOWLEDGEMENTS}
\medskip
\nobreak

We wish to thank Valerie Khoze, Michael Luke, Anatoly Radysuhkin, Martin
Savage, for many helpful conversations and suggestions on subjects related
to this work.  This work was begun while the authors attended the Aspen
Center for Physics and the Santa Fe Study Group.  We would like to thank
the organizers of those programs for their generous support.  L. R.
would like to thank the CERN Theory division for their hospitality
while this work was being completed.

\goodbreak
\medskip
\baselineskip = 18pt plus 1pt
\noindent{\bf FIGURE CAPTIONS}
\medskip
\nobreak
{\bf Figure 1:}
Virtual Compton scattering of a current coupled only to $h$-quarks
from a heavy meson, $H$, with momentum $P$.  The imaginary part is
proportional to the distribution function, $f(x,Q^2)$, at large-$Q^2$.  Only
the dominant process at large-$Q^2$ is shown, exclusive of QCD radiative
corrections.

\medskip
{\bf Figure 2:}
The cross section for an $h$-quark current to produce an $H$-meson
plus anything in the deep-inelastic limit.  Only the dominant contribution at
large-$Q^2$ is shown, exclusive of QCD radiative
corrections.  The cross
section is proportional to $\hat f(x,Q^2)$.

\goodbreak
\bigskip
\noindent{\bf REFERENCES}
\medskip
\nobreak
\item{1.}
B.~Adeva {\it et al.\/} (L3 Collaboration) {\it Phys. Lett.\/} {\bf B261}
(1991) 177; D.~Decamp {\it et al.\/} (ALEPH Collaboration) {\it Phys. Lett.\/}
{\bf B244} (1990) 551.
\medskip
\item{2.}
D.~Bortoletto {\it et al.\/} {\it Phys. Rev.\/} {\bf D37} (1988) 1719.
\medskip
\item{3.}
C.~Petersen {\it et al.\/} {\it Phys. Rev.\/} {\bf D27} (1983) 105.
\medskip
\item{4.}
V.~A.~Khoze, SLAC--PUB--5909, talk presented at the XXVI International
Conference on High Energy Physics, Dallas, 1992
\medskip
\item{5.}
B.~Mele and P.~Nason, {\it Phys. Lett.\/} {\bf B245} (1990) 635,
 {\it Nucl. Phys.\/} {\bf B361} (1991) 626.
\medskip
\item{6.}
R.~L.~Jaffe in {\it Relativistic Dynamics and Quark--Nuclear Physics\/},
M.~B.~Johnson and A.~Pickleshimer, eds. (Wiley Interscience, New York, 1986).
\medskip
\item{7.}
J.~C.~Collins and D.~E.~Soper, {\it Nucl. Phys.\/} {\bf B194} (1982) 445.
\medskip
\item{8.}
A.~Bassetto, M.~Ciafaloni and G.~Marchesini, {\it Phys. Rep.\/} {\bf 100}
(1983)
201.
\medskip
\item{9.}
M.~Luke, {\it Phys. Lett.\/} {\bf B252} (1990) 447.
\medskip
\item{10.}
R.~L.~Jaffe, in preparation.
\medskip
\item{11.}
R.~L.~Jaffe and M.~Soldate, {\it Phys. Rev.\/} {\bf D26} (1982) 49.
\medskip
\item{12.}
O.~Nachtmann, {\it Nucl. Phys.\/} {\bf B63} (1973) 237, {\bf B78} (1974) 455.
\nobreak

\bye